\documentclass[12pt]{article}
\usepackage{amsmath, amsfonts, amsthm, amssymb, color, graphicx, url, xr, zref, setspace, bm, algorithm}
\RequirePackage[colorlinks,citecolor=blue,urlcolor=blue]{hyperref}
\usepackage[top=2.8cm, bottom=2.8cm, left=2.54cm, right=2.54cm]{geometry}
\doublespace 

\author{
Steffen Ventz$^{\$}$\footnote{\small \small steffen.ventz.81@gmail.com}, Rahul Mazumder$^{\#}$, Lorenzo Trippa$^{\$}$ \\
{\small $^{\$}$Harvard T.H. Chan School of Public Health and Dana-Farber Cancer Institute}\\
{\small $^{\#}$Massachusetts Institute of Technology}
}

\usepackage[square,numbers]{natbib}
\begin{document}

\title{Integration of Survival Data from Multiple Studies}

{\small \date{\today} }
\maketitle

{\small
\begin{abstract} 
\noindent 	
We introduce a statistical procedure that integrates survival  data   from  multiple  biomedical studies,
to improve  the  accuracy  of  predictions of survival or other events,  based on individual clinical and genomic  profiles,
compared to  models developed  leveraging  only a single study  or    meta-analytic methods.
The method accounts for potential differences in  the relation between   predictors and outcomes across studies, 
due to  distinct  patient populations,  treatments and  technologies to  measure outcomes and biomarkers.
These differences are  modeled explicitly with study-specific parameters.  We use 
hierarchical regularization to  shrink the study-specific parameters  towards each other and to borrow information across studies.
Shrinkage  of the study-specific parameters 
is controlled by a similarity matrix, 
which   summarizes   differences and similarities of the relations between  covariates and outcomes  across studies. 
We illustrate the method in a simulation  study 
and using a collection  of gene-expression datasets in ovarian cancer. We show  
that the proposed model increases the accuracy of survival  prediction compared to alternative meta-analytic  methods. 
\end{abstract}
}


\section{Introduction}
Various biomedical  technologies enable the use of  omics  information  for  prognostic  purposes, 
to  quantify  the risk  of   diseases, or    to  predict  response  to treatments.
Risk  stratification  in oncology often utilizes  a set of biomarkers  to predict cancer progression 
      or  death   within  a   time   period. 
   The number of covariates  can exceed the  sample size.   
This makes     the identification of  relevant  genomic features for risk prediction  
 and  the development  of accurate    models challenging. 
Penalized regression  and 
methods that utilize multiple datasets have  been discussed  in this  context.  
Penalization methods    enable parameter 
estimation and prediction   when the number of predictors is large  \citep{Tibshirani1997}.   
 Meta-analyses \citep{dersimonian1986meta} and 
 integrated analyses \citep{conlon2006bayesian}
combine   information   from multiple studies for parameter estimation and prediction
\citep{hedges2014statistical, trippa2015bayesian}. 
These   statistical procedures  improve the estimation of parameters of interest with respect
  to single-study estimates if
 the  covariate-outcome  relations  are  similar across  studies \citep{riester2014, bernau2014cross, waldron2014comparative}.  
For instance \citet{riester2014, bernau2014cross} and  \citet{waldron2014comparative} showed that  meta-analytic  procedures 
tend to outperform   the  prediction accuracy of  models developed using only a single study.  
But \citep{riester2014, waldron2014comparative, trippa2015}  also described  heterogeneity of covariate effects across  cancer studies, due to differences in assays, treatments and  patient populations. 

We introduce a model for the integrated analysis  of a collection of  datasets, 
with the aim of improving  the accuracy  of  predictions, compared to single-study  models and  meta-analytic procedures.
%
We   use study-specific  parameters  in covariate-outcome regression models.   
These   parameters are  estimated 
borrowing   information across studies with hierarchical regularization,  which   
shrinks  the study-specific regression coefficients  towards each other.
%
We use a squared $K\times K$ similarity matrix   representative  of  differences and similarities of the covariates' effects  across  
$K$ studies. 
The matrix is  used to estimate the study-specific models. 
The regression parameters of  each study are shrunken more towards  the  parameters  
of   similar studies  and less  towards the remaining studies.  

In previous  work on integrative analyses  \citet{liu2011high}  
discussed  Bayesian  methods and variable selection for accelerated failure time models. 
Hierarchical normal models for multi-study  gene expression analyses have been developed in \citep{conlon2006bayesian, conlon2009hierarchical, conlon2012bayesian}, and
\citet{ma2011integrative, ma2011integrativeb} studied penalized regression methods for  integrative analyses, focusing on  binary and accelerated failure-time outcome models.
For case-control analyses   with multiple datasets,  \citet{liu2013estimation} proposed an adaptive group-LASSO (agLASSO) procedure,  and 
 \citet{Cheng2015} extended the  approach  using  different regularization techniques.
In the  following  sections  we  introduce    a      procedure  which builds  on the  work  that we mentioned.
The procedure shrink study-specific parameters towards  each  other  accounting for the  
  degrees  of  similarity  specific  of  each  pair  of  studies.

\section{The multi-study model}\label{Sec:Model}

We consider $K$ studies with  time-to-event   outcomes and predictors  such as gene expression measurements. 
For each study $k=1,\ldots,K$   the vector $\bm Y_k = \{ Y_{k,i}\}_{i=1}^{n_k}$ indicates  (possibly  censored) survival times of  $n_k$  individuals 
and $\bm C_k =\{C_{k,i} \}_{i=1}^{n_k}$ denotes the vector of censoring variables, where   
$C_{k,i}=1$  if $Y_{k,i}$ is an observed event time and it  is zero if there  is  censoring at time $Y_{k,i}$. 
The vector $\bm x_{k,i}  \in \mathbb R ^p$ represents a set of $p$ predictors 
and $\bm X_k = (\bm x_{k,1}, \cdots, \bm x_{k,n_k} )^T$. 
Lastly, $\mathcal D_k = ( \bm Y_k, \bm  C_k, \bm X_k )$  indicates the   data   from study $k$ and $\mathcal D = \{ \mathcal D_k \}_{k=1}^K$ is  a collection of studies. 

We  assume that  failure times in each study $k$ follow a proportional hazard model \citep{Cox1972} 
with baseline survival function $S_{k}(\cdot)$ and study-specific coefficients $\bm \beta_k \in \mathbb R^p.$ 
%
The 
approach that we will describe  can be applied to alternative time to event models, 
for instance  to accelerated failure time models \cite{Wei1992} or  accelerated hazards models  
\cite{chen2000analysis},
computations would only require minor modifications.  

Inference  is based on the
Breslow's modification  of  the partial log-likelihood functions \cite{Cox1972} for (possible tied) survival times,
$$  
l\left( \bm \beta_k, \mathcal D_k \right) 
=
\sum_{\ell= 1}^{m_k}  
\left\{   \bm { \tilde{x} }'_{k,\ell}  \bm  \beta_k  - d_{k,\ell}  \log \left ( \sum_{i : Y_{k,i} \geq t_{k,\ell} } \exp \{ \bm x_{k,i}'   \bm \beta_k \} \right)  \right  \},
$$
where $\big \{ t_{k,\ell} \big \}_{\ell=1}^{m_k}$ are the $m_k$ unique event times in study $k$, 
$d_{k,\ell}$ denotes the number of observed events at time $t_{k,\ell}$,
$\displaystyle \bm { \tilde{x} }_{k,\ell}  = \sum_{i} \bm x_{k,i}  I(C_{k,i} =1,  Y_{k,i} =t_{k,\ell} ),$ 
for $\ell = 1, \ldots, m_k,$  
and $I(A)$ is the indicator function of the event $A$.

When the number of predictors  exceeds the number of observed events 
 a unique maximum partial-likelihood estimate  does not exist and
maximization of    the regularized likelihood function 
$ l\left( \bm \beta_k, \mathcal D_k \right) -  R\left( \bm \beta_k \right)$
%
has been proposed  \citep{Tibshirani1997, Park2007, Simon2011} to obtain covariate effect estimates.  
Here    $R\left( \bm \beta_k \right)$ is a non-negative function that equals zero when $\bm \beta_k = \bm0$.
Popular    approaches include 
the LASSO, ridge,  elastic-net and the bridge penalties
to name  a few \citep{Hoerl1970, Tibshirani1996, Tibshirani1997, fu1998penalized, Wang2008, Simon2011}. 
We refer to \cite{bovelstad2007predicting, bovelstad2009survival, van2009survival, bernau2014cross} for  comparisons   of  regularization methods  for predicting survival outcomes using  genomic  profiles.  

As  demonstrated  in \citep{bernau2014cross} 
 studies, with nearly  identical aims, often presents different joint  distributions  of  predictors  $\bm x_{k,i}$  and  outcomes $Y_{k,i}$.
Clusters  of  studies which correspond to  different essays, 
patient populations,
treatments, and study designs have  been discussed \citep{bernau2014cross, trippa2015}. 
We introduce a  model with study-specific  parameters 
$\bm \beta_k$, 
and a  latent parameter $\bm \beta_0$, which can be interpreted as  the  mean parameter  across studies.
Some studies will have similar   vectors  $\bm \beta_k$ due to similarities in the assays  and patient populations, 
while other studies might be  considerably  different  \citep{bernau2014cross, waldron2014comparative}.
We estimate the vectors $\bm \beta_k$  by borrowing information from studies $k'\neq k$ that are similar to study $k$. 
At the same time,  
studies $k'$ that differ substantially from study $k$ will have  little  influence on the estimation of   $\bm \beta_k$.
In different words, borrowing  of information   mirrors  similarities and differences  across   studies. 
The latent parameter and study-level  parameters 
 $\bm \beta=( \bm \beta_0,  \cdots, \bm \beta_K )$ are estimated  using the regularized likelihood 
\begin{align}\label{HierarchicalRegularization}
l_R\left( \bm\beta    \right) = 
 \sum_{k=1}^K  l\left( \bm \beta_k, \mathcal D_k \right) 
-R_0 (\bm \beta_0) -R_1\left( \bm \beta  \right).
\end{align}

%
Here the parameters  $\bm \beta_k$ can be interpreted as a noisy realization of  $\bm \beta_0$, the average effect across studies.
The  non-negative function $R_0(\cdot)$ regularizes  
  $\bm \beta_0$  and is zero when $\bm \beta_0=\bm0$ 
(for example a lasso penalty). 
Similarly, the  non-negative function $R_1(\cdot)$  is zero  
when $\bm \beta_0 =   \bm \beta_1 = \cdots =  \bm \beta_K$
(see below for examples) 
and is used to borrow  information across studies in the estimation of  $\bm \beta$.
 In our applications in Sections \ref{Sec:Simulation:Study} and \ref{Sec:Ovarian:Cancer} 
we will use $\widehat{\bm \beta}_0$ for risk predictions of patients in populations $k>K$ that  are not  represented in our  collection of $K$ studies, whereas for patients belonging to populations $k=1,\dots, K,$  the  estimate $\widehat{\bm \beta}_k$ can be directly  used for risk predictions.

Penalized maximum   likelihood estimates
based on  (\ref{HierarchicalRegularization})   have a  Bayesian interpretation.  
See \citep{Tibshirani1996, hastie2015statistical, polson2015proximal, park2008bayesian} for  a  discussion on the  relations  between  regularization methods   and  Bayesian analyses. Consider a Bayesian model for  the unknown parameters $\bm \beta$, 
with prior probability 
$Pr(\bm \beta_0) \propto e^{ -R_0 (\bm \beta_0) }$ for the vector $\bm \beta_0$ 
and 
$Pr(\bm \beta_1, \dots, \bm \beta_K  | \bm \beta_0)\propto e^{ -R_1 \left( \bm \beta  \right)   }$ 
for 
the study specific parameters  conditionally on  $\bm \beta_0$. 
The  approximate posterior density of $\bm \beta$ with respect to  the partial likelihood 
(see \cite{Sinha2003} for a formal justification) is  proportional to  
%
\begin{align}\label{BayesianModel1}
Pr_{PL}  \left( \bm \beta | \mathcal D \right)
&\propto Pr(\bm \beta_0)   Pr(\bm \beta_1, \cdots,  \bm \beta_K | \bm \beta_0)   \prod_{k=1}^K e^{ l(\bm \beta_k, \mathcal D_k)}.
\end{align}
%
Therefore the mode of  (\ref{BayesianModel1})  coincides  with  
the parameter $\bm \beta$  that maximizes  (\ref{HierarchicalRegularization}).
If  we  set $R_1\left( \bm \beta  \right) = \sum_{k} \tilde{R}_1 \left( \bm \beta_k,  \bm \beta_0  \right)$, with 
$ \tilde{R}_1 \left( \bm \beta_k,  \bm \beta_0  \right) \geq 0$ 
then the    Bayesian model (\ref{BayesianModel1}) 
incorporates  the
assumption that
studies are exchangeable with, conditionally on $\bm \beta_0$,  independent and  identically distributed covariate effects $\bm \beta_k$. 
For example     
$\tilde{R}_1 \left( \bm \beta_k, \bm \beta_0  \right) =  || \bm \beta_k - \bm \beta_0  ||^2_2/(2\lambda_1)$ 
and 
$R_0\left(    \bm \beta_0 \right) =  ||   \bm \beta_0  ||_2^2/(2\lambda_0)$   
is  consistent  with the commonly utilized hierarchical normal  prior model  with, {\it  a priori}, 
  correlations 
$\mbox{Cor}( \beta_{k,j},  \beta_{k',j} ) = \lambda_0/(\lambda_0 +\lambda_1 )> 0$
for  studies $k' \neq k$  when the latent vector $\bm \beta_0$ is integrated out.
This  regularization implies 
positive and symmetric borrowing of information for all pairs  $k\neq k'$ of studies, and
may not be appropriate for  groups of studies with different patient populations.

For the latent mean  parameter $\bm \beta_0$  we use  the elastic-net penalty \citep{zou2005regularization}, 
$$R_0( \bm \beta_0 ) =  \lambda_0 ||\bm \beta_0||_1  + \lambda_1  ||\bm \beta_0||_2^2,$$ 
$\lambda_0,\lambda_1\ge0,$ with LASSO
and ridge penalty as special cases, when $\lambda_1=0$ and $\lambda_0=0$ respectively. 

To account for   differences  and similarities of the  available  studies,     
we use 

\begin{align}\label{Pooling::Studies}
R_1 \left( \bm \beta \right) 
 =  \sum_{j=1}^p 
 ||  \bm \beta_{1:K,j} -  \beta_{0,j} \bm 1  ||_{ \bm \Sigma} ^{a}
\end{align}  
in (\ref{HierarchicalRegularization}),  
where  
$ \bm 1$  is a K-dimensional vector with one on each component,  $\bm \beta_{1:K,j} =(\beta_{1,j}, \cdots, \beta_{K,j} )'$
  refers  to   covariate $j=1, \cdots, p$  in each study, and
$|| \bm x||_{\bm \Sigma} = \sqrt{ \bm x' \bm \Sigma^{-1} \bm x}$.
The symmetric matrix $\bm \Sigma$
is  positive-semidefinite and enables differential borrowing 
of information across studies. 

For $a=2$, the  minimizer of  (\ref{HierarchicalRegularization})   is equivalent to  the posterior mode  
  when,   {\it  a priori},
     the coefficients  $\bm \beta_{1:K,j}, j=1, \cdots, p$ across studies  are modeled  
  as  multivariate  normal   with mean  $\beta_{0,j} \bm 1$ and covariance matrix $\bm \Sigma $. 
In this case  $\Sigma_{k,k'} = 0$  implies that  
 $\bm \beta_k$ and  $\bm \beta_{k'}$ are, 
{\it  a priori} and conditionally on $\bm \beta_{0}$ independent.
Whereas a large covariance $\Sigma_{k,k'} > 0$  indicates similarities between  $\bm \beta_k$ and  $\bm \beta_{k'}$.

For $a=1$ the penalty $R_1(\bm \beta)$ in (\ref{Pooling::Studies})  becomes   the sum of  Mahalanobis distance of  
$\bm \beta_{1:K,j}$
from the mean  $\beta_{0,j} \bm 1$ 
with covariance matrix   $  \bm \Sigma$.   
With   $\bm \Sigma  \propto \bm I$ this penalty reduces  to the  group LASSO 
\citep{Cheng2015, liu2013estimation} 
with one group for each covariate $j=1, \cdots, J$.   


The regularization parameters $ \lambda_0, \lambda_1, a, $ and $\bm \Sigma$ 
determine 
(i) the sparsity of   $\widehat{\bm \beta}_0$  (the number of components $\widehat{ \beta}_{0,j}=0$)
and 
(ii) the similarity of the  estimates $\widehat \beta_{1,j}, \cdots, \widehat \beta_{K,j}$  across studies, 
including the number of  identical  study-specific estimates $\widehat \beta_{k,j}=  \widehat \beta_{k',j}$.

When $a\geq 1$, $\bm \Sigma$ is positive-definite  and $ \lambda_0 > 0$  or $ \lambda_1 > 0$,
the regularized log-partial-likelihood  (\ref{HierarchicalRegularization}) is  concave.    
If we fix  $\lambda_1\geq 0,$  $a\geq 1$  and the positive-definite matrix $\bm \Sigma,$  then
the number of    components $\widehat{\beta}_{0,j} $  equal to $0$ increases with $\lambda_0.$ 
For instance, consider
 $a>1$ in (\ref{HierarchicalRegularization}), $\lambda_1=0$,
and 
$g(\bm \beta_0) =  \max_{ \bm \beta_1, \cdots, \bm \beta_K  }  \sum_{k=1}^K  l\left( \bm \beta_k, \mathcal D_k \right) 
-R_1\left( \bm \beta  \right).$
The   concave map $g(\bm \beta_0)- \lambda_0  || \bm \beta_0||_1$   bounds  the regularized log-partial-likelihood.  
If   we choose $\lambda_0 $ larger  than $\max_{1 \leq j\leq p} | \partial g(\bm \beta_0) / \partial  \beta_{0,j} | $ at $\bm \beta_0 = \bm 0$, then (\ref{HierarchicalRegularization}) is maximized at $\widehat {\bm \beta_0}=\bm 0$.
In contrast,  for values $\lambda_0$ below this maximum 
some of the  estimates $\widehat {\beta}_{0,j}$  will be different from zero.

For $\lambda_0, \lambda_1 \geq 0,$ 
and $a=1$, the  choice of  
$\bm \Sigma$  can  
lead to 
identical study specific estimates $\widehat \beta_{1,j}= \cdots = \widehat \beta_{K,j}$. 
%
We provide an example with $\lambda_0=0 $ and $ \bm \Sigma = \sigma^2 \bm I$.
Let  $\bm z_k = \bm \beta_k -\bm\beta_0, k=1, \cdots, K$, and define
$h(\bm z) = \max_{\bm \beta_0} \sum_{k=1}^K  l\left( \bm z_k + \bm \beta_0, \mathcal D_k \right)  -R_0\left( \bm \beta_0  \right).$
  The map   $h(\bm z)- \sum_{1\leq j\leq p} ||  (z_{1,j},\ldots, z_{K,j} )  ||_{ \bm \Sigma} ^{a}$  bounds the re-parametrized regularized log-partial-likelihood ($l_R: [\bm \beta_0, \bm z_1, \cdots, \bm z_K]\rightarrow \mathbb{R}$).
If we specify $1/ \sigma > \max_{j,k} |\partial h( \bm z) / \partial  z_{k,j} |$  at   $\bm z = \bm 0$,  
then  the concave function (\ref{HierarchicalRegularization}) is maximized  at $\widehat {\bm \beta_1}=\ldots=\widehat {\bm \beta_K}=\widehat {\bm \beta_0}$.
More generally, if we don't  assume  a diagonal  $\bm \Sigma$ and  indicate
with $\sigma^2$  the largest eigenvalue of $\bm \Sigma$,  then  the equalities $ \widehat {\bm \beta_k}=\widehat {\bm \beta_0}$
hold when $1/ \sigma > \max_{j,k} |\partial h( \bm z) / \partial  z_{k,j} | $  at   $\bm z = \bm 0$.


\section{Parameter estimation}
We use an alternating direction method of multipliers algorithm \citep{boyd2011} 
to estimate  $\bm \beta$, see  \citep{boyd2011}  for an introduction to this  algorithm.
We first formulate the optimization of (\ref{HierarchicalRegularization}) with respect to 
$\bm \beta= (\bm \beta_0, \cdots, \bm \beta_K )$
as a constrained convex minization problem 
$$\min_{ (\bm \beta, \bm z) }  \Bigg  \{  \sum_k -l(\bm \beta_k, \mathcal D_k) + R_0(\bm \beta_0)  + R_1(\bm z) \Bigg \},$$
where $\bm z =  (\bm z_0, \cdots, \bm z_K )', \bm z_k \in \mathbb R^p$,
subjected to the affine constraints
$\bm \beta_k = \bm z_k, k=0, \ldots, K$.
We then introduce for this minimization problem the scaled augmented Lagrangian
\begin{align}\label{Multipliers}
L_\rho (\bm z,  \bm \beta, \bm u) 
=
\sum_{k=1}^K   - l(\bm \beta_k, \mathcal D_k) + R_0(\bm \beta_0) + R_1(\bm z) + \sum_{k=0}^K  \frac{ \rho} { 2}  || \bm \beta_k - \bm z_k + \bm u_k ||^2_2,
\end{align}
where  $\rho>0,$ with augmented $\bm u = (\bm u_0, \cdots, \bm u_K), \bm u_k \in \mathbb R^p$.
For a fixed  $\rho>0,$ the  algorithm  that we describe
 converges to a solution   $\bm \beta - \bm z =  \bm u = \bm 0$
 that  maximizes (\ref{HierarchicalRegularization}).  
The algorithm  minimizes (\ref{Multipliers}) iteratively 
(i)  with respect to $\bm \beta$, and  
(ii) with respect to $\bm z$, 
and 
(iii) then it updates  $\bm u$ to $ \bm u \leftarrow \bm u + \bm \beta- \bm z$,
while keeping at each of the  three  steps the remaining two  parameters fixed.
At each iteration of the algorithm the minimization of (\ref{Multipliers}) 
 with respect to $\bm \beta$  ({\it step i})  can be carried out independently for each component $\bm \beta_k$, $k=0,\ldots,K$, 
and the minimization with respect to $\bm z$    ({\it step ii}) can be carried out  independently by covariates $j=1, \cdots, p$.

The algorithm starts with  an initial estimate of  $\bm \beta$   
(we use $\bm 0$ or  preliminary  estimates of $\bm \beta_k, k=0, \cdots, K$), 
$\bm \beta = \bm z$
and $\bm u = \bm 0$. 
At each iteration,  in {\it step i}  the algorithm 
minimizes  (\ref{Multipliers})  $\bm \beta,$ keeping  $\bm z$ and $\bm u$  fixed,
by  setting $\bm \beta_{0}= \dfrac{ S\big( \rho (\bm z_{0} - \bm u_{0}),  \lambda_0\big)} { \rho + 2 \lambda_1}$
where $S(\bm x, \lambda)$ is the
coordinate-wise soft-thresholding function $s(x_j, \lambda)=(1-\lambda /|x_j|)_{+} x_j$  \citep{Tibshirani1996},  
and 
$$\bm \beta_{k} = \arg \min_{\bm b} \Big ( -l(\bm b, \mathcal D_k) + 
 \rho  || \bm b -   \bm z_k + \bm u_k   ||_2^2 /2 \Big)$$
   for the remaining  $k=1, \cdots, K$. 
We used a low-memory 
 quasi-Newton algorithm \citep{byrd1995} for the latter minimization $k>0$. 

In {\it step ii},  the  algorithm    minimizes   (\ref{Multipliers}) 
with respect to $ \bm z$ keeping  $\bm \beta$ and $\bm u$ fixed. 
This is done independently for each  covariate $1 \leq j \leq p$, 
because $R_1(\cdot)$  and the $l^2_2$-norm in  (\ref{Multipliers}) 
can be factors into the sum of $p$ terms each involving only the $j$-th row of $\bm z=(\bm z_0, \cdots, \bm z_K)$  and
the $j$-th row of  $\bm \beta + \bm u.$ 
For example,
when  $a=2$ in (\ref{Pooling::Studies}),  
$\bm z' = \Big( \bm I +  2 \bm H' \bm \Sigma^{-1} \bm H /\rho  \Big)^{-1} (\bm \beta + \bm u)' ,$ 
where the $K$ by $K+1$ matrix   $\bm H = [-\bm 1, \bm I ]$ 
is the concatenation of  $-\bm 1$ with k-dimensional  identity matrix $\bm I$.  
This, computation is implemented by first computing the matrix $K+1$ by $K+1$ matrix $ \Big( \bm I +  2 \bm H' \bm \Sigma^{-1} \bm H /\rho  \Big)^{-1}$,  and then multiplying it with each column $j=1, \cdots, p$  of  $(\bm \beta + \bm u)'.$

Lastly,  in {\it step iii}, 
 $\bm u$ from the last iteration is updated to $\bm u +   \bm \beta - \bm z.$  
We iterate this three steps until 
the  $l^2_2$-norm of both $\bm z - \bm \beta$ and the difference between $\bm z$ from two successive iterations becomes smaller than
a  pre-specified treshold $\epsilon > 0$  \citep{boyd2011}.

\section{Simulation Study}\label{Sec:Simulation:Study}

We consider a total of $18$ studies. 
Either $2,5,10$ or $15$ of these 18 studies are used to estimate the model (\ref{HierarchicalRegularization}).  
The remaining $16, 13, 8$ or $3$ studies are used for out-of study evaluations.
For each study $k=1, \cdots, 18$, we drew the sample size $n_k$ of the study from a uniform distribution $n_k \sim Unif(100, \cdots, 500),$
and then generated the covariates $x_{k,i}\in \mathbb R^{500}$ of observations $i=1, \cdots, n_k$ from a normal distribution $\bm x_{k,i} \sim N_{500}(\bm 0, \bm V) $  with covariance $V_{j,j'} = 0.3^{|j-j'|}$ between variables $j$ and $j'$. 

\begin{figure}[h]
\includegraphics[scale=.67]{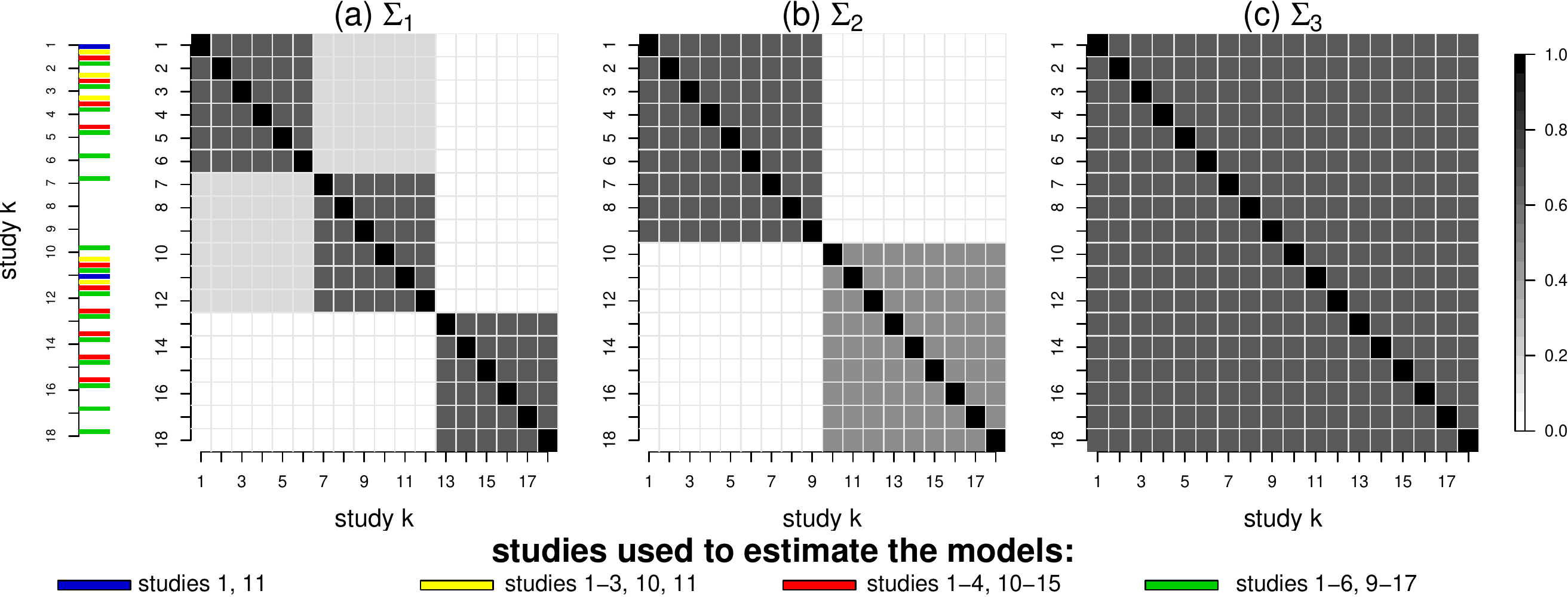}
\caption{Similarity matrixes $\bm \Sigma=\bm \Sigma_1,\bm \Sigma_2, \bm \Sigma_3  \in \mathbb R^{18 \times 18}$ used to simulate the datasets. 
We simulated collections of $18$ datasets $\mathcal D =\{D_k\}_{k=1}^{18}$ with similarity matrix $\bm \Sigma$ for $\bm \beta.$ 
Studies  indicated in blue (2 studies), yellow (5 studies), red (10 studies) or green (15 studies) are used to fit models with $K=2, 5, 10$ or $15$ studies. 
}\label{Fig:TrueCor}
\end{figure}

We then generated 100 times the parameters $\bm \beta \in \mathbb R^{ 500 \times 19 }$ and 
a collection of 18 studies $\mathcal D = (\mathcal D_k)_{k=1}^{18}$.
In each of these 100 simulations we first generated the vector   $\bm \beta_0 \in \mathbb R^{500}$ 
from a two-component mixture distribution with 
(i) a point mass at zero and 
(ii) a normal distribution with  mean zero and variance $0.1$.  
The  proportion of zeros of  this mixture distribution equals $p_0 = 0.9$ or $0$.
We then generated independently $p=500$ vectors  
$(\epsilon_{1,j},\ldots,\epsilon_{K,j}) \sim N_K(\bm 0, \bm \Sigma)$  
and set $\beta_{k,j} = \beta_{0,j} + \epsilon_{k,j}$ for each covariate $j=1,\ldots,p$ and study $k=1,\ldots,K$.
We consider three   matrices 
$\bm \Sigma = \bm \Sigma_1, \bm \Sigma_2, \bm \Sigma_3$ (see Figure \ref{Fig:TrueCor})
with 3, 2 or a single cluster of studies.

Survival times where generated from  proportional hazard models 
with baseline survival functions $\widehat{S}_k(\cdot)$,  
regression coefficients $\bm \beta_k$, 
and censoring survival functions $\widehat S_{C,k}(\cdot)$. 
Here $\widehat{S}_k(\cdot)$ and $\widehat S_{C,k}(\cdot)$ have been estimated  from  the ovarian cancer datasets
that  we discuss in Section \ref{Sec:Ovarian:Cancer}.
For each study $k$ we also generated an additional 1,000 observations,
that were    not used to fit regression models, but were used to evaluate predictions.

\subsection{Estimation of $\Sigma$ and  selection of $(\lambda_1, \lambda_0)$}\label{Sec:Estimate:Similarity}

We use  initial estimates $\widehat{ \bm \beta_k}$ 
obtained from $K$ independent ridge regression models to estimate $\bm  \Sigma$. 
The procedure  leverage  the Bayesian interpretation   (\ref{BayesianModel1}) of the
regularized likelihood (\ref{HierarchicalRegularization}).
As formalized in  (\ref{BayesianModel1}), 
with $R_1 \left( \bm \beta \right) 
 =  \sum_{j=1}^p 
 ||  \bm \beta_{1:K,j} -  \beta_{0,j} \bm 1  ||^a_{ \bm \Sigma} ,$ 
we can interpret   
$(\beta_{j,1}, \cdots, \beta_{j,K}), j=1, \cdots, p$, 
as $p$ independent  vectors each with  covariance matrix $\bm \Sigma$.
If the  $\bm \beta_k, k=1, \cdots, K,$ were known we could  straightforwardly estimate  $\bm \Sigma$.  
For instance with $a=2$, the parameters $(\beta_{j,1}, \cdots, \beta_{j,K}), j=1, \cdots, p$
 can be interpreted as 
independent multivariate normal vectors with mean zero and covariance matrix $\bm \Sigma$.  The joint normal  distribution implies  that
$E[ \bm \beta_k | \{ \bm  \beta_{k'} \}_{0<k'\le K, k'\neq k}   ] =   \sum_{ 0<k'\le K, k'\neq k}   \alpha_{k,k'} \bm  \beta_{k'}$
where the weight vector $\bm \alpha_{k}=(\alpha_{k,k'})_{0<k'\le K, k'\neq k}$ is as function of $\bm \Sigma$ \citep{eaton1983multivariate} 
for  each $k=1, \cdots, K$.
Therefore  the conditional 
expectation of  $\bm X_k  \bm \beta_k$, given $\{ \bm  \beta_{k'} \}_{0<k'\le K, k'\neq k}$  
is  $\sum_{ 0<k'\le K, k'\neq k}   \alpha_{k,k'} ( \bm X_{k} \bm  \beta_{k'}).$ 
After replacing $\bm  \beta_{k'}$ with  out  initial estimates $\widehat{ \bm \beta}_{k'}$, 
we  estimate  $\bm \alpha_{k}$ 
   via   a Cox model with $K-1$  covariates
 $z_{k'}=\bm X_{k} \widehat{ \bm \beta}_{k'}$ and  regression coefficients
 $\bm \alpha_{k}$.
We then use the  empirical covariance matrix of  
$\bm \beta_k^\star=\sum_{0<k'\le K, k'\neq k}  \widehat{\alpha}_{k,k'} \widehat{ \bm \beta_k}, k=1, \cdots, K$  as an estimate of $\bm \Sigma.$
Note that $\bm \beta_k^\star$  has  a  direct interpretation  under the  assumption  that  the  independent  vectors $(\beta_{1,j},\ldots,\beta_{K,j})$
    are  in a  linear subspace  with  dimension less than $K$.

\begin{figure}[h]
\begin{center}
\includegraphics[width=15cm, height=10cm]{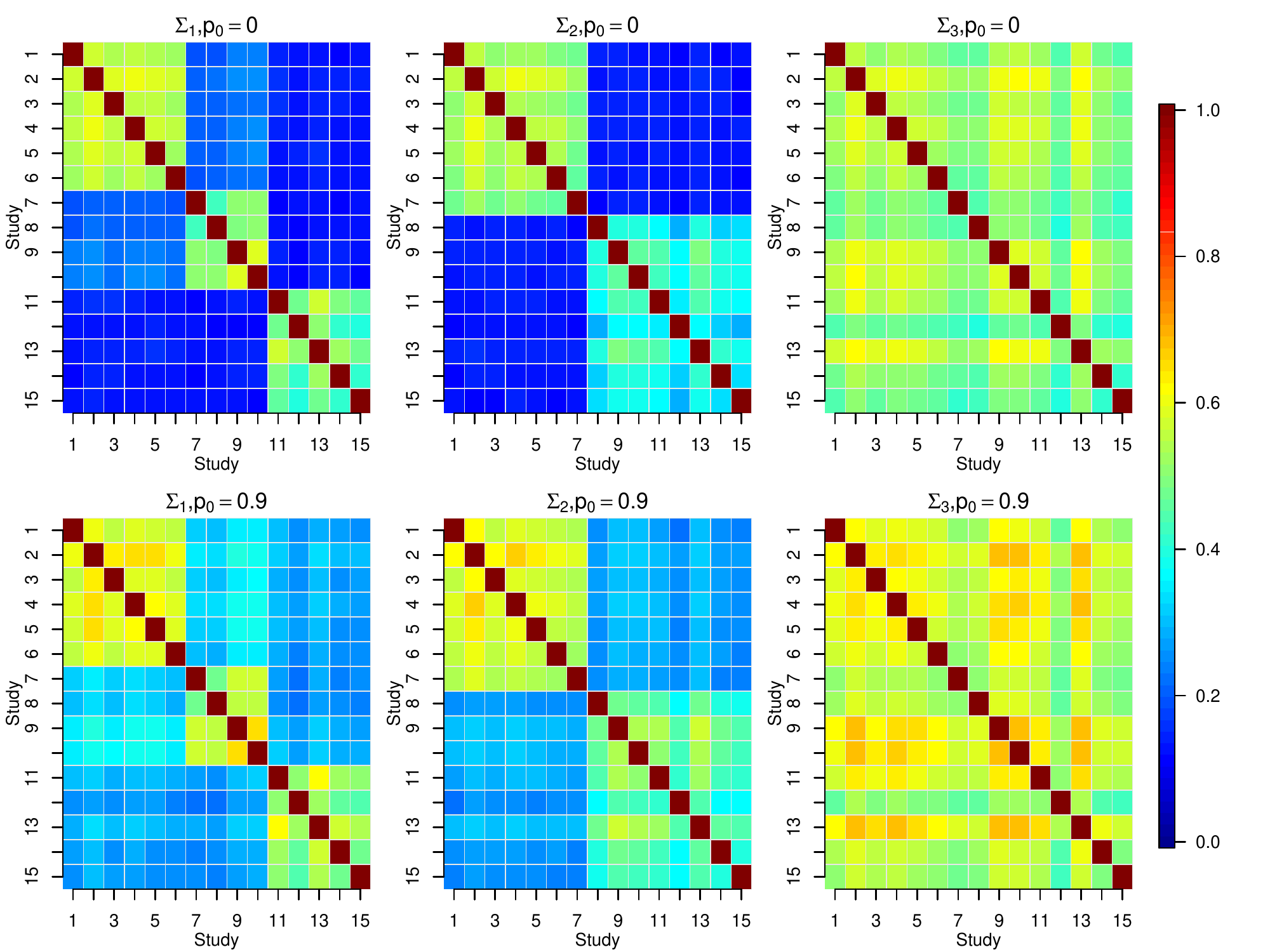}
\caption{Average estimates across 100 simulations of the similarity  matrix of the regression coefficients between the $K=15$ studies. 
}
\label{Fig:CorEstimate}
\end{center}
\end{figure}

Figure \ref{Fig:CorEstimate} 
shows averages across the 100 simulations of the estimated similarity matrix between studies for  
the largest model with $K=15$ studies 
when  $p_0=0$ (top row) and $p_0 = 0.9$ (bottom row). 
Figures  \ref{Fig:CorEstimate} and  \ref{Fig:TrueCor} show that the algorithm of Section \ref{Sec:Estimate:Similarity}  on average recovers the similarity structure of the 15 studies.

To select the parameters $\lambda_0$ and/or $\lambda_1,$ 
we use   Monte-Carlo cross-validation (CV) \citep{shao1993linear}.
We  evaluate candidate parameter estimates  $\widehat{\bm \beta}$ using   
$
\mathcal C(\widehat{\bm \beta}) = \sum_{k}   w_k \mathcal C( \widehat{\bm \beta}_k, \mathcal D_{k}),$ 
where the C-statistics
$\mathcal C( \widehat{\bm \beta}_k,   \mathcal D_k ) = 
 \widehat{Pr} \Big( \bm x_1' \widehat{\bm \beta}_k  > \bm  x'_2 \widehat{\bm \beta}_k  | Y_1 < Y_2  \Big)$  
is the estimated concordance \citep{harrell1984regression, pencina2004overall, uno2011c}
between two survival times $Y_1$ and $Y_2$ with covariate vectors $\bm x_1$ and $\bm x_2$ in population $k$. 
The weights $w_k\geq 0$ account for differences in study sample sizes, 
we used $w_k = 1/\sqrt{m_k}$.
We first split the data randomly $M$-times into  training (80\%) and validation  (20\%) datasets. 
Next we define a grid of tuning parameters $\bm \lambda = (\lambda_1, \lambda_0).$
For each combination of   tuning parameters $\bm \lambda$  of the grid, 
we estimate  $\widehat {\bm \beta}^{(m)}$ based on the $m=1, \cdots, M$   CV training datasets (which are identical across different grid-points) and use the validation 
datasets to obtain estimates of the study-specific C-statistics 
$\mathcal C_\lambda(\widehat {\bm \beta}_k^{(m)}, \mathcal D_{k}),  m=1, \cdots, M$ for $\widehat {\bm \beta}_k^{(m)}$ with $\bm \lambda$. 
We then average these $M$ C-statistics and compute the overall estimate $\mathcal C_\lambda(\widehat {\bm \beta})$ for $\lambda$.
Lastly, we select the $\lambda$-value with  the  highest average  C-statistics $\mathcal C_\lambda(\widehat {\bm \beta}).$

\subsection{ Prediction Accuracy }
\begin{figure}[h]
\begin{center}
\centering 
\includegraphics[height=11cm, width=17cm]{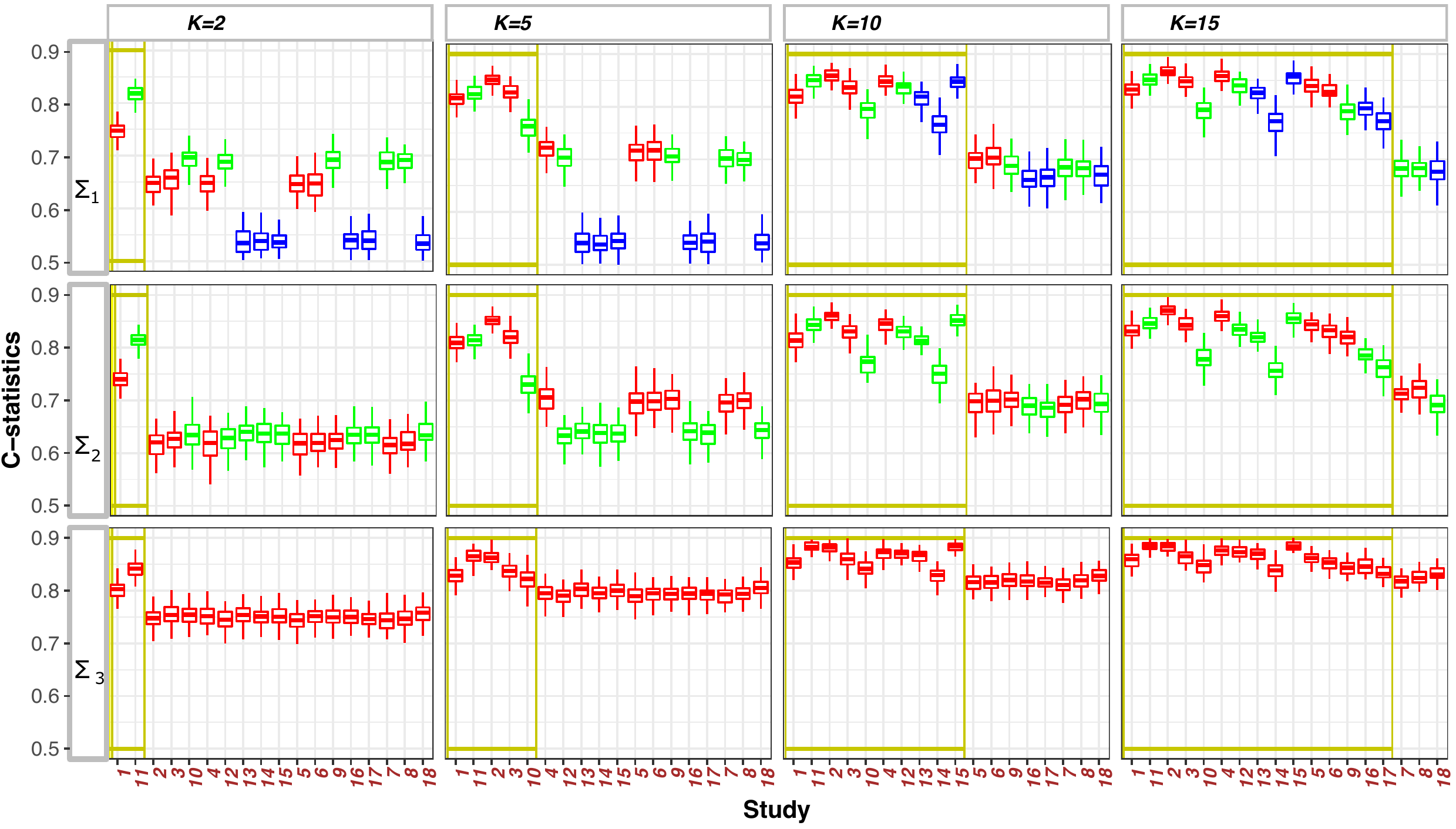}
\caption{
Predictions with  the penalized regression model with $(a,\lambda_0)=(2,0)$. 
We consider $\bm \Sigma= \bm \Sigma_1, \bm \Sigma_2$ and $\bm \Sigma_3$ (see Figure \ref{Fig:TrueCor}), and  $p_0=0$  across 100 simulations of a collection of 18 studies. 
Either $K=2,5, 10$ or $15$ studies (studies inside the brown rectangles) 
were used for estimation/selection of $(\bm \Sigma, \lambda_1, \bm \beta)$.  
The colors (red, green, blue) of the Box-plots indicate clusters of  studies  (3, 2, or 1 clusters when  $\bm \Sigma=\bm \Sigma_1$, $\bm \Sigma_2$ or $\bm \Sigma_3$). 
}\label{Fig:Learning:Accross:Studies}
\end{center}
\end{figure}

Figure \ref{Fig:Learning:Accross:Studies} shows, for each of the 18 studies 
box-plots of the estimated C-statistics \citep{harrell1984regression, pencina2004overall, uno2011c} when  either $2, 5, 10$ or $15$  studies (1st to 4th column) were used to estimate the similarity matrix and the model.  
C-statistics $\mathcal C(\widehat{\bm \beta}_k, \mathcal D_k)$ for studies $k$ that were utilized  to estimate the model are highlighted inside the brown rectangles (estimated using the additional 1000
hold-out   observations  in study $k$), 
whereas  C-statistics $\mathcal C(\widehat{\bm \beta}_0, \mathcal D_k)$ for studies $k$ that were not used to estimate the model
are shown on the right of the  brown rectangles. 

The three rows of Figure \ref{Fig:Learning:Accross:Studies} correspond 
 to scenarios with data generated using $\bm \Sigma_1$ (top row of Figure \ref{Fig:Learning:Accross:Studies}), $\bm \Sigma_2,$  (2nd  row), or $\bm \Sigma_3$ (bottom row)  as  illustrated in Figure \ref{Fig:TrueCor}. 
Red, green and blue  box-plots on the top-row indicate the three clusters of studies under $\bm \Sigma_1$. Similarly, red and green box-plots in the 2-nd row indicate the two clusters of studies under $\Sigma_2.$  
Differences in the distribution of the C-statistics between studies within the same cluster are due 
to differences in the sample sizes  $n_k$ and covariate matrixes $\bm X_k$, 
which remain identical across the simulated datasets.

For $\bm \Sigma = \bm \Sigma_1$ (1st row of Figure \ref{Fig:Learning:Accross:Studies}),  
with three clusters of studies, predictions   
show  improvements when the number of studies used to train the regression models  increases $K$. 
For $K=2$ or $5$,
all studies  used for estimation  belong to the first two clusters (red and green box-plots).
In these two  cases, for each  study $k=13, \cdots18$ in cluster 3 (blue box-plots) 
the inter-quartile range (IQR) of the C-statistics 
$\mathcal C(\widehat{\bm \beta}_0, \mathcal D_k)$ across simulations lies within the interval 0.52 to 0.55. 
%
Whereas for $K=10$ (3rd column, studies 1-4 and 10-15 are use for estimation), 
studies from all three clusters have been used for training.  
In this case, the IQRs of $\mathcal C(\widehat{\bm \beta}_0, \mathcal D_k)$ across simulations 
for all three hold-out studies $k=16, 17, 18$ in cluster 3 are within the interval 0.65 to 0.69.
The  last row of Figure  \ref{Fig:Learning:Accross:Studies} shows that, as expected,
borrowing of information in the estimation of model parameters 
is most effective in the case of a single cluster of studies.

\medskip

Next, we compared our  estimates of $\bm \beta$ based on model (\ref{HierarchicalRegularization}),
with    
$a=2$ for  $R_1(\cdot)$ and  ridge penalty (HR-R, $\lambda_1=0$) or LASSO penalty (HR-L, $\lambda_2=0$) for   $\bm \beta_0,$
to Cox models trained separately on each study $\mathcal D_k$ with LASSO (single-study LASSO, SL) or ridge penalties (single-study ridge,  SR) for $\bm \beta_k$. 
%
In addition we consider two models that combine all (2, 5, 10 or 15) studies  into a single dataset 
and estimate a single Cox model (with regression parameters $\bm \beta_0$)  
using a LASSO (pooled LASSO, PL) or ridge (pooled ridge, PR) penalty for the  coefficients    $\bm \beta_0.$
We also consider  two  meta-analysis approaches   described in \citep{waldron2014comparative, riester2014}  
that  combine study specific estimates $\widehat{\bm \beta}_k$ into a single   vector $\widehat{\bm \beta}_0$
using either fixed effects (FE) or random-effects  (RE) estimation.

\begin{figure}[p]
\begin{center}
\centering 
\includegraphics[scale=0.7]{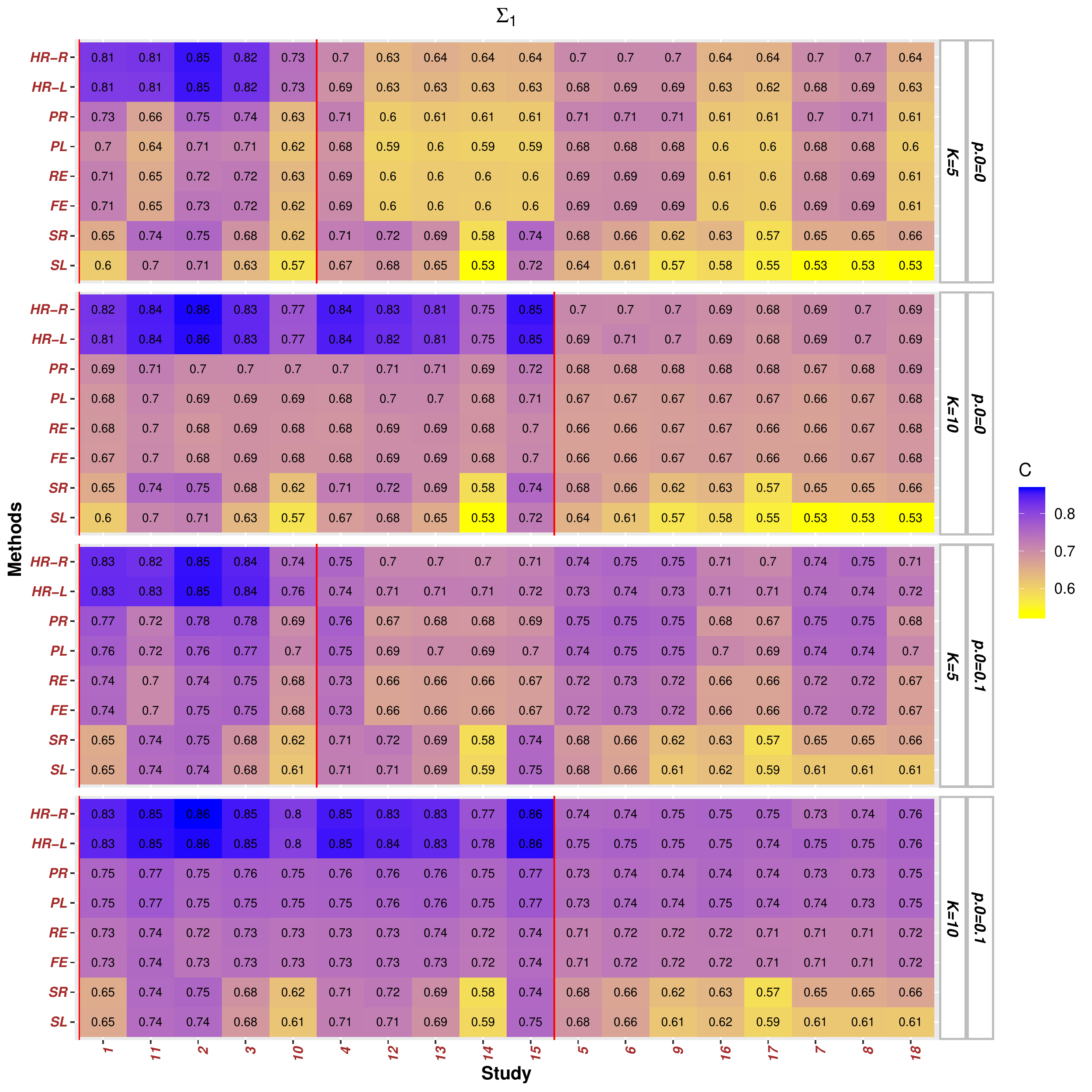}
\caption{ 
Average prediction across 100 simulations of  a collection of 18 studies. 
Study specific effects $\bm \beta_k$ have been generated under $\bm \Sigma_1$ (see Figure \ref{Fig:TrueCor}). 
Either $5$ or $10$ of the 18 studies (studies in the left of the horizontal red bar) 
are used for the similarity matrix and covariate-effect estimation. 
See Figures \ref{SFig:Prediction2} and \ref{SFig:Prediction3}
for results with similarity scenarios $\bm \Sigma_2$ and $\bm \Sigma_3$.
}\label{Fig:Prediction1}
\end{center}
\end{figure}

Figures \ref{Fig:Prediction1} and supplementary Figures  \ref{SFig:Prediction2} and   \ref{SFig:Prediction3} 
show the average C-statistics of each method when we used  $K=5$ or $K=10$ studies for  estimation. 
%
The pooled LASSO and ridge models (PL and PR) and 
the  meta-analyses methods  (FE and RE) estimate a single parameter
$\bm \beta_0$, which was  used  to compute the C-statistics  
$C(\widehat{\bm \beta}_0, \mathcal D_{k} )$ 
for each study $k$.
For the single-study SL and SR models 
we used the study-specific estimates  $\widehat{\bm \beta}_k$ to 
compute  $C(\widehat{\bm \beta}_k, \mathcal D_{k} )$ for in-study prediction (using the 1,000  validation observations).
For prediction with SL and SR in studies $k'$ not used for estimation, 
we used each estimate  $\widehat{\bm \beta}_k$ of the $K=5$ (or 10) training studies
for predictions $\mathcal C(\widehat{\bm \beta}^k, \mathcal D_{k'} )$ in all hold-out studies $k'$. 
For each hold-out studies $k'$ we then  averaged  these    $\mathcal C(\widehat{\bm \beta}^k, \mathcal D_{k'} )$ over all $K=5$ (or 10) training studies, 
i.e.  Figure \ref{Fig:Prediction1} reportes  $\sum_{k} \mathcal C(\widehat{\bm \beta}^k, \mathcal D_{k'} )/K$  for studied $k'$.

For studies $k$ used to train the model, both
HR-L and HR-R improve predictions $C(\widehat{\bm \beta}_k, \mathcal D_{k} )$ substantially
compared to single-study estimates SL and SR.  
For instance, with $K=5, p_0=0$ and unknown $\bm \Sigma=\bm \Sigma_1$ (three clusters of studies),    
the average difference between $C(\widehat{\bm \beta}_k, \mathcal D_{k})$ of HR-R and SR 
is between  0.07 and 0.16 for each of the five studies 
(0.62 to 0.75 for SR compared to 0.73 to 0.85 for HR-R). 
Similarly, meta-analytic and pooled estimates FE, RE and PL, 
SL improve predictions on the $K=5$ datasets  compared to 
single-study estimates, especially PR. 
But improvements are smaller than for  HR-P and HR-L, with 
 C-values on average (across simulations) between 0.08 to 0.15 below HR-R and HR-L models 
 (for instance, 0.63 to 0.73 for PR compared to 0.73 to 0.85 for HR-R).
When $K=10$ studies are used to estimate the models, results are similar to the setting with $K=5$ - meta-analytic, pooled and hierarchical estimates improve predictions over single-study estimates, 
with larger improvements for HR-R and HR-L estimates for in-study predictions.    

In the case of a single cluster of studies 
($\bm \Sigma=\bm \Sigma_3$, supplementary Figure \ref{SFig:Prediction3}), 
with strong similarity of the study-specific parameters  $\bm \beta_k$, 
pooling of studies to estimate a single $\bm \beta_0$
is expected to be the most favorable prediction approach.
Therefore,  PL, PR, HR-R and HR-L predict survival substantially 
better than the   FE, RE, SL and SR methods
(supplementary Figure \ref{SFig:Prediction3}). 
With $p_0=0$, in-study predictions based on  HR-R and HR-L estimates  are  
on average slightly better than for PR and PL estimates 
(difference of 0.01 to 0.04 for HR-R compared to PR with  $K=5$ studies, 
and 0.02 to 0.05 with $K=10$).  
Whereas PR, PL, HR-R and HR-L have similar average $C$-statistics for holdout studies.   

\section{Survival prediction in ovarian cancer}\label{Sec:Ovarian:Cancer}

We applied   model (\ref{HierarchicalRegularization})
to predict
survival  in ovarian cancer 
using the {\it curatedOvarianData} repository, a curated collection of gene-expression datasets  
\citep{ganzfried2013curatedovariandata}.
To evaluate  prediction, we split  the largest study in the database, 
the TCGA dataset \citep{cancer2011integrated} with $510$ observations,  
1,000 times randomly into a training dataset of  $n_{1}=50, 75, \ldots,$ or $300$  observations 
and a validation dataset with $510 - n_{1}$ observations.
We predicted patient   survival  $Y_{1,i}$ in the TCGA holdout data 
by leveraging the hierarchical regularization model (\ref{HierarchicalRegularization})
using five additional datasets $k=2, \cdots, K=6$
(PMID-17290060\citep{dressman2007integrated}, GSE51088\citep{karlan2014postn}, MTAB386\citep{bentink2012angiogenic}, GSE13876\citep{crijns2009survival} and GSE19829 \citep{konstantinopoulos2010gene})
with  sample sizes ranging between  $n_k = 42$ (GSE19829) and $157$ (GSE13876) observations. 
In all the  analyses we used the expression values  of the $p=3,030$ genes 
that are common in all six studies 
to predict  patient   survival.
%

To evaluate the hierarchical regularization method (\ref{HierarchicalRegularization}),  
we created different cross-study heterogeneity scenarios that are motivated by documented inconsistencies across cancer  datasets and    
by possible pre-processing  errors   \citep{NDHHS2015, potti2006genomic, bonnefoi2007retracted, acharya2008gene, potti2006gene, salter2008integrated}. 
This is achieved by  introducing in one (scenario 2: GSE13876\citep{crijns2009survival}) or two  studies (scenario 3: GSE13876\citep{crijns2009survival} and GSE19829 \citep{konstantinopoulos2010gene})  
 a distortion  
of the expression values $x_{k,i,j}$ which become $10-3 x_{k,i,j}, j=1, \ldots, p$ for study $k=K$ (scenario two) or studies $k=K-1, K$ (scenario three).  
In scenario one  we used the covariates $x_{k,i,j}$ of the six studies. 

Similar to Section 4, we consider 
parameter estimates  based on the 
$n_1=50, \cdots, 300$  TCGA training samples using 
 (i) single study Cox models with LASSO (SL) or 
 (ii) ridge (SR) regularization, 
pooled Cox regression models  
that combine the  $n_1$ TCGA-observations  and the remaining five  studies (PMID-17290060, GSE51088, MTAB386, GSE13876 and GSE19829) 
into a single dataset with  
(iii)  LASSO (PL) or 
(iv)  ridge (PR) regularization,
(v)   fixed effects  (FE) and 
(vi)  random effects (RE) model meta-analyses models as described in  
\cite{riester2014, bernau2014cross, waldron2014comparative}, 
and (vii) the proposed hierarchical regularization model  (\ref{HierarchicalRegularization}) with $\lambda _0=1, a=2$ (HR-R).

Single-study cox-models with LASSO-penalty trained on the TCGA data with $n_1=50$  
data points had low average C-values of $0.505$ across the 1,000 generated training-validation samples,  
with minor improvements up $0.52$ when  $n_1=300$  
observations are used for model training. 
Single study ridge regression models performed substantially better, 
with average C-statistics ranging between $0.53$ for $n_1=50$ 
and $0.57$ for $n_1=300$ observations.  
Improvements in risk predictions through integration of additional studies 
vary substantially across data-integration methods and scenarios. 
For scenario 1, 
FE  and RE meta-analyses have both nearly constant and identical average C-statistics of $0.57$ 
across all sample sizes $n_1$,  while PL had an average  C-statistics of  $0.56$ for $n_1=50$
with minor improvements up to $0.57$ when  $n_1=300.$
%
Both, HR-R and PR have similar prediction accuracy across sample sizes $n_1,$
with identical average C-statistics  of $0.60$ when $n_1=50$ and 
modest improvements up to $0.61$ for both, HR-R and PR, when $n_1=300$.
%

\begin{figure}[h]
\begin{center}
\centering 
\includegraphics[scale=0.7]{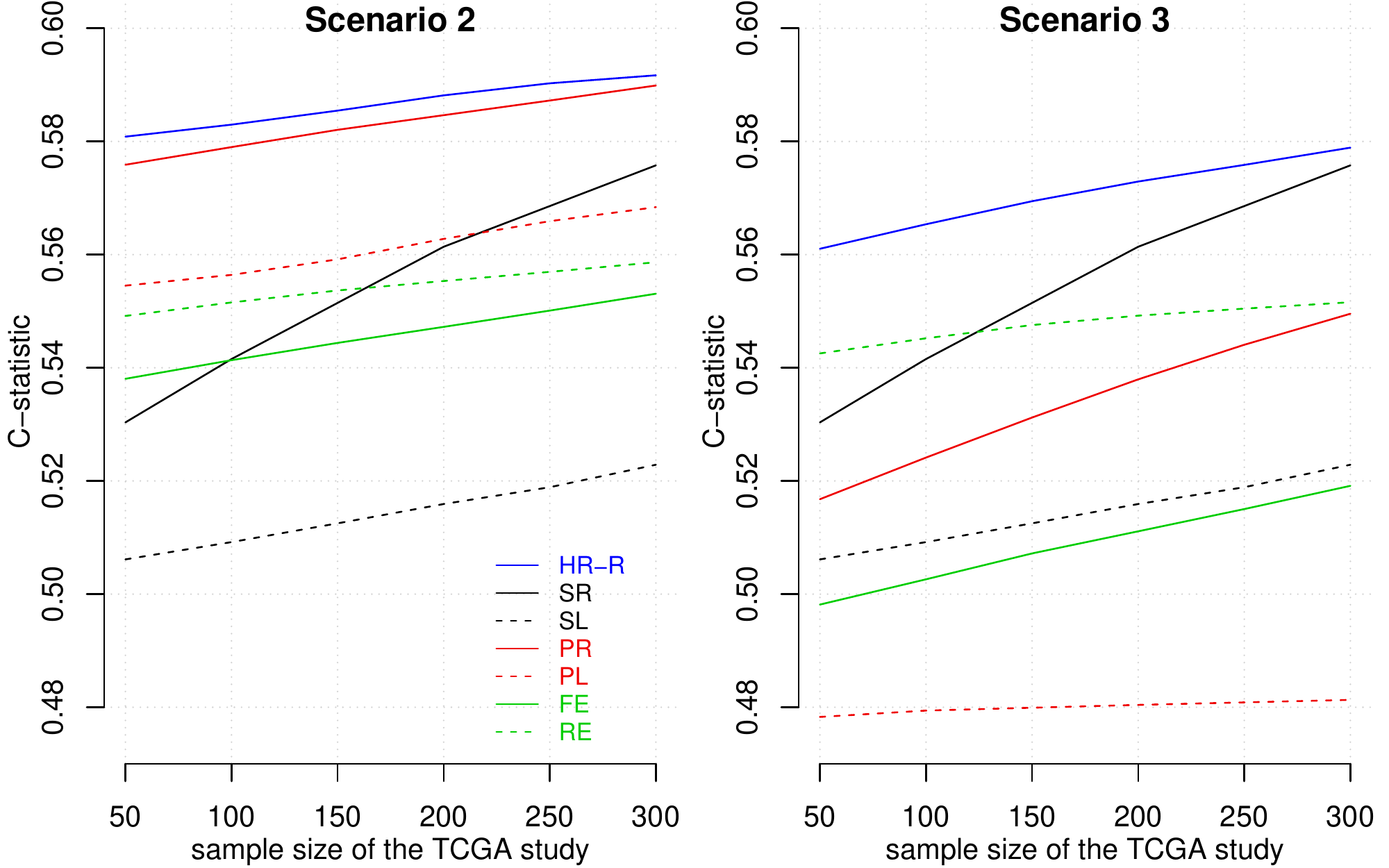}
\caption{ 
Average C-statistics for single-study SL and SR methods with $n_1=50, \cdots, 300$ TCGA training samples, and for data-integration methods (PL, PR, FE, RE, HR-R) use  the $n_1$ TCGA training samples and  training samples from five additional  studies (PMID-17290060, GSE51088, MTAB386, GSE13876 and GSE19829). 
}\label{Fig:OC}
\end{center}
\end{figure}

Figure \ref{Fig:OC} shows, for scenarios two and three, 
average C-statistics for  the TCGA validation samples. 
Different curves correspond to different prediction methods.
The black curves show the average C-statistics  (y-axis)
across  the 1,000  TCGA validation samples of  size $510-n_1$  
for  Cox models trained  on  $n_1=50, \cdots, 300$  observations from  the TCGA study  (x-axis)  
using either  SL (dotted curve) or  SR (solid curve).
The red curves show the average C-statistic 
for PR (solid  curve) and PL (dotted  curve) models, 
the green curves correspond to  FE (dotted line) and RE (solid line) meta-analysis models, 
and the blue curve corresponds to the HR-R.

In scenario two, the
RE meta-analysis, which   combines estimates from the $n_1=50$  TCGA data points with estimates from the remaining five studies,  
has the same  average C-statistics as the single-study  SR model trained on $n_1=240$ patients.    
For  sample sizes $n_1>250$,
pooled regression models PR have similar performances as RE models.
The    HR-R model trained on $n_1=50$ TCGA patients 
has an average C-statistics the is superior to those of PR and FE procedures
with  $n_1 = 50, \cdots, 300$. 
As expected, 
 with  increased discrepancies  in  the  relations  between covariates and  outcomes across studies
 (scenario three), 
performances of all data-integration methods decrease. 
HR-R 
 models with  $n_1=50$ TCGA patients  
 have similar average C-values as single study SR modes with 
$n_1 \approx 200$ patients.   %
PR, PL, FE and RE methods rely on the assumption that the regression parameters are similar across studies. With substantial departures from this assumption the hierarchical model HR-R  shows, across all sample sizes $50 \leq n_1 \leq 300$,
 gains in average prediction accuracy compared to PR, PL, FE and RE.

\section{Discussion}\label{Sec:Discussion}

The analysis of  relations between  omics variables and time to event outcomes, 
and the use  of individual profiles $\bm x_{k,i}$ for predictions,  are  particularly challenging  
when the sample size $n_k$ is small. These analyses often  include thousands of potential predictors.
The use of  multiple studies and  pooling of  information 
can  improve prediction accuracy. 
 Meta-analyses can be utilized when the  relations  of covariates and outcomes are homogeneous across studies. 
But recent  work 
 in oncology  \citep{riester2014, waldron2014comparative, trippa2015}   showed   
that there can be  clusters of  studies  with relevant  discrepancies   in their covariate-outcome relations  due,  for  example, 
to differences in study designs, patient populations and treatments. 

We combined two  established concepts, 
regularization of  regression models \citep{Hoerl1970, Tibshirani1996, Tibshirani1997, fu1998penalized} 
and  metrics of similarity between datasets \citep{hastie2009elements} that identify clusters of studies.  
We used these concepts   
to estimate  study-specific  regression  parameters  $\bm \beta_{k}$ and  for  predictions, both 
in  $k=1,\ldots,K$ contexts that are represented in  our  collection of datasets, 
for example  $K$  distinct  geographic regions,
and  in  other  contexts  $(k=K+1)$ by  estimating the latent   parameters $\bm \beta_{0}$.

The $K\times K$ similarity matrix $\bm \Sigma$ is used to regularize  the likelihood function, and it 
 tunes  the degree of borrowing of information in the estimation of $K$ study-specific regression models. 
It shrinks the estimate of the k-th study-specific regression parameter $\bm \beta_{k}$ towards estimates $\bm \beta_{k'}$ of  studies $k'$ that are similar to study $k$ (large $\Sigma_{k,k'}$).
In contrast studies with low similarity ($\Sigma_{k,k'} \approx 0$) have little  influence  on the estimation of  $\bm\beta_{k}$.
 In our analyses we verified  that,  if there are  clusters  of studies with similar  predictors-outcome relations,    then
the introduced  method
       improves the accuracy of predictions   compared to alternative  procedures, including  single-study estimates, meta-analyses and  pooling of all studies into a single data matrix.

\subsubsection*{Funding}
Lorenzo Trippa  was  partially supported by the NSF grant 1810829.

\bibliography{Lit.bib}  

\clearpage

\appendix
\renewcommand\thefigure{\thesection.\arabic{figure}}    
\setcounter{figure}{0}

\section{Appendix}
\begin{figure}[h]
\includegraphics[scale=0.7]{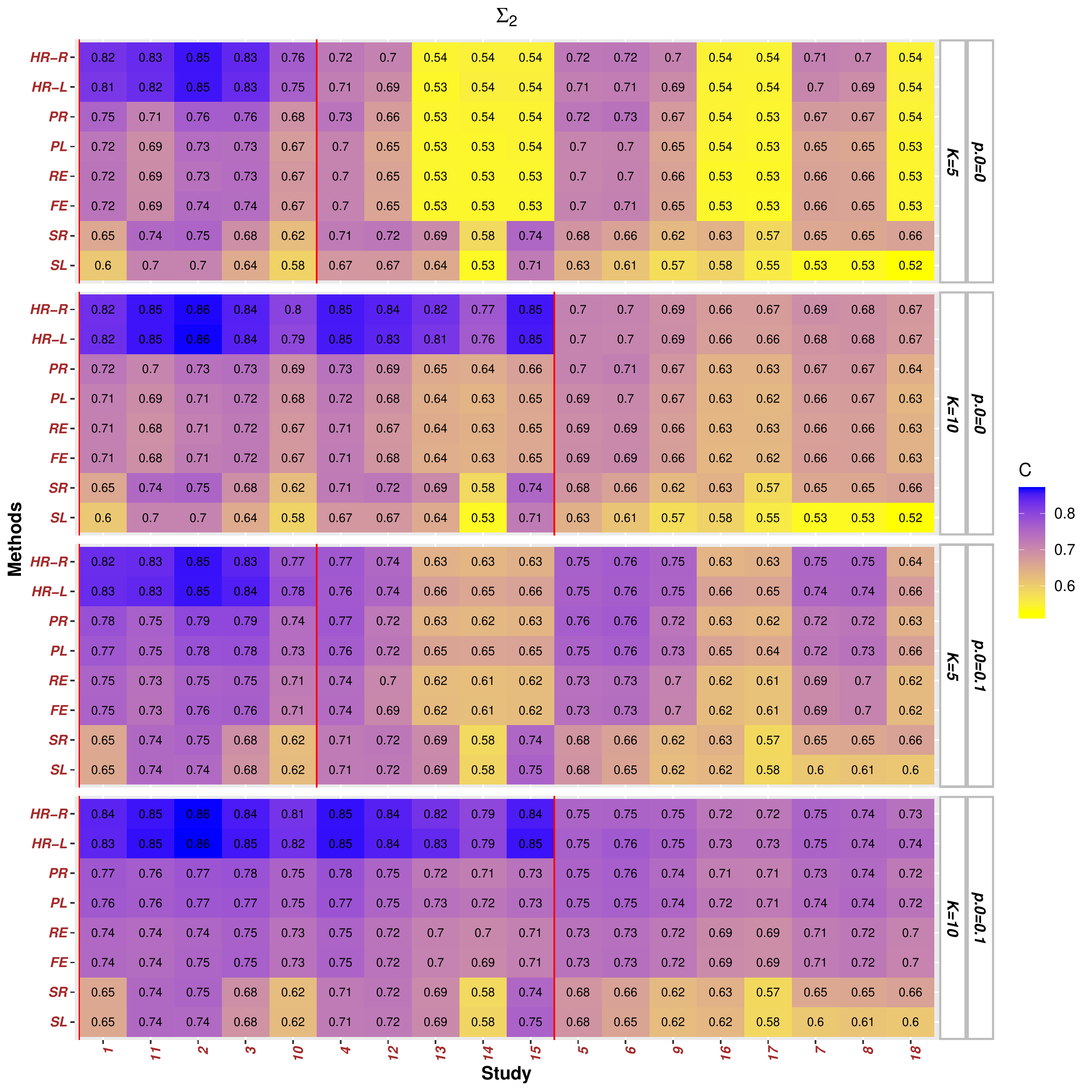}
\caption{ 
Average prediction across 100 simulations of  a collection of 18 studies. 
Study specific effects $\bm \beta_k$ have been generated under $\bm \Sigma_2.$ 
Either $5$ or $10$ of the 18 studies (studies in the left of the horizontal red bar) 
are used for the similarity matrix and covariate-effect estimation. 
}\label{SFig:Prediction2}
\end{figure}

\begin{figure}[h]
\includegraphics[scale=0.7]{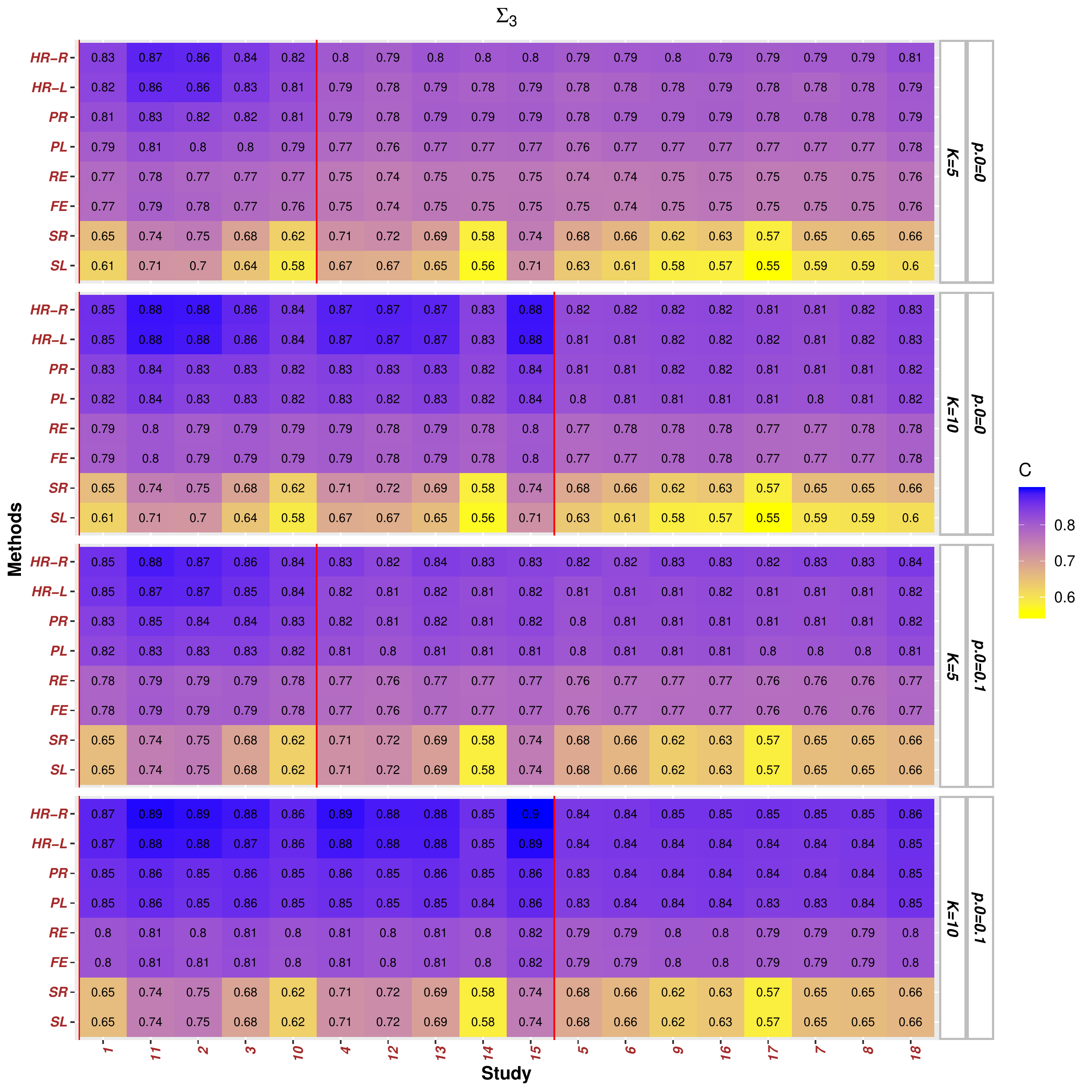}
\caption{ 
Average prediction across 100 simulations of  a collection of 18 studies. 
Study specific effects $\bm \beta_k$ have been generated under $\bm \Sigma_3.$ 
Either $5$ or $10$ of the 18 studies (studies in the left of the horizontal red bar) 
are used for the similarity matrix and covariate-effect estimation. 
}\label{SFig:Prediction3}
\end{figure}

\end{document}